\documentclass[a4paper,12pt]{article}

\usepackage[margin=0.75in]{geometry}
\usepackage{amsmath}
\usepackage{amssymb}
\usepackage{multicol}
\usepackage{bigints}
\usepackage{tikz-network}
\usepackage{tcolorbox}
\usepackage{setspace}
\usepackage{enumitem}
\usepackage{graphicx}
\usepackage{mathtools}
\usepackage{physics}
\usepackage{amsthm}
\usepackage{amsmath,amssymb,xcolor}
\usepackage{graphics}
\usepackage{adjustbox}
\usepackage{booktabs}
\usepackage{subcaption}

\usepackage{txfonts}
\usepackage{multirow}
\usepackage{hyperref}
\usepackage[square,numbers]{natbib}
\title{On Novel Approach for Computing Distance based Indices of Anti-tuberculosis Drugs}

\usepackage{authblk}
\author[1*]{D.C. Gunawardhana}
\author[2]{ G.H.J. Lanel}
\author[3]{ K.K.K.R. Perera}
\author[4]{ A.G.M.J. Gunarathna}
\affil[1]{Department of Mathematical Sciences, Faculty of Applied Sciences, South Eastern University of Sri Lanka}

\affil[2]{Department of Mathematics, Faculty of Applied Sciences, University of Sri Jayewardenapura, Sri Lanka}

\affil[3]{Department of Mathematics, Faculty of Science, University of Kelaniya, Sri Lanka}

\affil[4]{Department of Chemistry, Faculty of Science, University of Kelaniya, Sri Lanka}

\usepackage{longtable}
\usepackage[english]{babel}

\newtheorem{definition}{Definition}

\providecommand{\keywords}[1]
{
	\small	
	\textbf{\textit{Keywords---}} #1 
}

\begin{document}
	
	\maketitle
	\begin{abstract}
		
		This work aims to assess the molecular architectures of anti-tuberculosis drugs using both  degree-based topological indices and novel distance based indices. We can represent the chemical arrangement as a graph, with atoms serving as the vertices and connections as the edges. Here, the multi bonds were considered as multi edges and included all the hydrogen atoms. Also, we consider three dimensional molecular graph. As a result, the actual bond lengths have been used for computation of new distance based indices. Compared to numerous studies, this is a significant improvement.  Furthermore, the investigation of these indices includes a study on the quantitative structure-property relationship (QSPR). The research demonstrates a notable correlation between these indicators and the physical attributes of anti-tuberculosis drugs. Here, Since we reduced the some of existing critical assumptions in the literature, chemists and pharmaceutical professionals might potentially eliminate the need for clinical trials by employing this theoretical model. These models would enable them to predict the characteristics of anti-tuberculosis medications.
	\end{abstract}
	
	\footnote{* Corresponding author.\\
		Email addresses: rehan@seu.ac.lk\ (D.C. Gunawardhana), ghjlanel@sjp.ac.lk\ (Prof. G.H.J. Lanel),\ kkkrperera@kln.ac.lk (Dr. K.K.K.R. Perera), medhagunaratna@kln.ac.lk(Dr. A.G.M.J. Gunaratna) }
	
	\section{Introduction}
	

	Tuberculosis(TB) is a bacterial infection caused by \textit{Mycobacterium tuberculosis}, leading to primarily lung damage. The primary site of TB infection is the lungs, although it can also impact other organs such as the brain, spine and kidneys. TB continues to be a significant global health issue, well-known for its notorious bacterial infection. TB is transmitted into the air when individuals with lung TB cough, sneeze, or spit upon another. An individual just need inhalation of a small number of germs to become sick. WHO says that, annually, a staggering 10 million individuals become afflicted with tuberculosis (TB). Not with standing its preventability and curability, tuberculosis claims the lives of 1.5 million individuals annually, therefore establishing itself as the leading infectious cause of mortality worldwide \cite{dis17}. Although most individuals who contract tuberculosis reside in low- and middle-income nations, the disease is prevalent worldwide. A majority of individuals afflicted with tuberculosis are concentrated in eight nations, namely Bangladesh, China, India, Indonesia, Nigeria, Pakistan, Philippines, and South Africa \cite{dis17}.
	
	  Despite the significant reduction in TB-related deaths ascribed to the advent of antibiotics, the investigation of the molecular structures of infections and potential treatments has become more vital due to the rise of antibiotic-resistant genotypes.
	
	Hence, the examination of molecular graphs has acquired considerable significance in the field of cheminformatics, including chemistry, mathematics, and information science. An in-depth understanding of the molecular organization of a chemical compound is crucial in the field of chemistry. A molecular graph is a visual representation of the chemical formula of a molecule, with atoms displayed as vertices and chemical bonds as edges. The computation of topological indices that quantify different properties of molecular structures is crucially facilitated by the graphical depiction of these structures.
	
	Numerical values derived from the molecular graph can be used to depict the structural properties of a molecule. The aforementioned techniques have been widely utilized in studies concerning Quantitative Structure-Activity Relationships (QSAR) and Quantitative Structure-Property Relationships (QSPR). Such investigations are essential for predicting the biological activities and physicochemical properties of chemical substances. The Wiener index, developed by Harry Wiener in 1947, was the initial topological statistic to demonstrate a pronounced correlation with the boiling points of alkane compounds \cite{dis11}. Further studies have shown a relationship between the Wiener index and several properties of molecules, including critical point constants \cite{dis12}, density, surface tension, viscosity \cite{dis13}, and van der Waals surface area \cite{dis14}.
	
	Despite the proven utility of topological indices in predicting certain properties, it is important to recognize that no single indication can accurately correlate with all the physical attributes of a chemical entity. This limitation emphasizes the need of doing additional study on new topological indices that can provide more comprehensive forecasting capabilities in QSAR/QSPR studies.
	
	Object of this work is to assess the advancement of topological indices in enhancing the accuracy of modeling chemical and biological properties of compounds. Specifically, the project will examine their application in addressing antibiotic resistance and enhancing the efficiency of drug discovery. Thus, this work makes a substantial contribution to the field of cheminformatics by greatly enhancing the knowledge of molecular topology. These findings have far-reaching implications in scientific disciplines such as pharmacology, toxicology, and materials research.
	
	The graph $G(V, E)$ with the set of vertices $V(G)$ and the set of edges $E(G)$ is connected if there exists a connection between any pair of vertices in $G$. The graph theoretical distance between two vertices $u$ and $v$ is denoted as $d(u, v) = d_G(u, v)$ and is defined as the length of shortest path between $u$ and $v$ in graph $G$. The number of vertices of $G$ adjacent to a given vertex $v$ is the “degree” of this vertex and will be denoted by $d_v(G)$. In all most all the studies, researchers have used two dimensional($2D$) molecular graph and \(d(u,v)\), the graph theoretical distance. For molecular graphs, \(d(u,v)=1\). Then it is impossible to get meaningful results by considering the graph theoretical lengths. Therefore, we propose to use the actual bond length of three dimensional($3D$) molecular structure of relevant medicine instead of \(d(u,v)\). When it comes to $3D$ molecular graph, let $bl(u,v)$ denotes the geometrical distance between two adjacent atoms.

In \cite{dis15}, M. Adanan et al.  have computed the degree-based indices for anti-tuberculosis drugs without considering multi bonds and omitting hydrogen atoms.	Similarly, in the almost all studies, researchers have omitted the \(C-H\) bond and double/ tipple bonds purposefully to get the required results. Due to the their approaches, the important information which provide from the real structure might be missed. In our work, we consider the full molecule structure to compute the required results.

	Some of the degree-based topological indices which we consider in this study are deﬁned as follows.
	\begin{definition}
		$ABC$ index \cite{dis1} is proposed by Estrada et al. defined as, \[ ABC(G)=\displaystyle\sum\limits_{uv \in E(G)}\sqrt{\dfrac{d(u)+d(v)-2}{d(u)d(v)}}\]
	\end{definition}
	\begin{definition}
		The Randic index is proposed by Milan Randic \cite{dis2}, as
		
		\[R(G)=\sum\limits_{uv \in E(G)}\sqrt{\dfrac{1}{d(u)d(v)}}.\]
	\end{definition}
	\begin{definition}
		The sum connectivity index \cite{dis3} is proposed by Zhou and Trinjstic in, as \[S(G)=\sum\limits_{uv \in E(G)}\sqrt{\dfrac{1}{d(u)+d(v)}}.\]
	\end{definition}
	
	\begin{definition}
		The GA index is proposed by Vukicevic et al. \cite{dis4} in \[GA(G)=\sum\limits_{uv\in E(G)}\dfrac{2\sqrt{d(u)d(v)}}{d(u)+d(v)}.\]
	\end{definition}
	\begin{definition}
		The f\/irst and second Zagreb indices \cite{dis5} are proposed by Gutman and Trinajestic, as
		
		\[M_1(G)=\sum\limits_{uv \in E(G)}d(u)+d(v)\]
		\[M_2(G)=\sum\limits_{uv \in E(G)}d(u)d(v).\]
	\end{definition}
	
	\begin{definition}
		The Harmonic index \cite{dis6} is proposed by Fajtlowiczet al. as, \[H(G)=\sum\limits_{uv \in E(G)}\dfrac{2}{d(u)+d(v)}\]
	\end{definition}
	\begin{definition}
		The hyper-Zagreb index \cite{dis7} is proposed by Shirdelet al. as,
		\[HM(G)=\sum\limits_{uv\in E(G)}\left( d(u)+d(v)\right)^2 \]
	\end{definition}
	
	\begin{definition}
		The third Zagreb index \cite{dis8} is proposed by Fath-Tabar et al. as \[ZG_3(G)=\sum\limits_{uv\in E(G)}\left|d(u)-d(v) \right|. \]
	\end{definition}
	
	\begin{definition}
		The forgotten index \cite{dis9} is proposed by Furtula et al. as, \[F(G)=\sum\limits_{uv\in E(G)}d^2(u)+d^2(v).\]
	\end{definition}
	
	\begin{definition}
		The symmetric division index \cite{dis10} is proposed in as, \[SSD(G)=\sum\limits_{uv\in E(G)}\left( \dfrac{P}{Q}+\dfrac{Q}{P}\right) \]
		where, \(P=\min\left\lbrace d(u),d(v)\right\rbrace \), \(Q=\max\left\lbrace d(u),d(v) \right\rbrace \).
		
	\end{definition}
	\begin{definition}
		The Wiener index \cite{dis16} is denoted by $W(G)$ and defined as,
		\[W(G)=\dfrac{1}{2}\sum\limits_{u,v \in V(G)} d(u,v).\]
	\end{definition}
	
	\begin{definition}
In 1995, M.J. Klein et al. introduced the Hyper Wiener index \cite{dis18} denoted by $WW(G)$ and  defined as,
		\[WW(G)=\dfrac{1}{2}\left( \sum\limits_{u,v \in V(G)} d(u,v)+\sum\limits_{u,v \in V(G)} {d(u,v)}^2\right) .\]
	\end{definition}
	
	\begin{definition}
	 Plav\v si\'c et al.\cite{dis19} and Ivanciuc et al. \cite{dis20} independently introduced	the Harary index denoted by $Ha(G)$ and defined as,
		\[Ha(G)=\sum\limits_{u,v \in V(G)} \dfrac{1}{d(u,v)}.\]
	\end{definition}
	
	\begin{definition}
	In order to characterize alkanes by an integer, in 1989, Schultz introduced , the Schultz index \cite{dis21} denoted by $S\!ch(G)$ and defined as,
		\[S\!ch(G)=\sum\limits_{u,v \in V(G)} (d_G(u)+d_G(v))d(u,v).\]
	\end{definition}
	
	\begin{definition}
In 1996,	S. Klavžar and I. Gutman introduced	Modified Schultz index/ Gutman index \cite{dis22} is denoted by $Gut(G)$ and is defined as,
		\[Gut(G)=\sum\limits_{u,v \in V(G)} (d_G(u)\cdot d_G(v))d(u,v).\]
	\end{definition}
	
	\begin{definition}
In1982,	Alexandru T. Balaban \cite{dis23, dis24} introduced the Balaban index denoted by $J(G)$ and  defined as,
		\[J(G)=\dfrac{m}{m-n+2}\sum\limits_{u,v \in V(G)} \dfrac{1}{d(u,v)+1}.\]
	\end{definition}
	
	\begin{definition}
		The Klein index is denoted by $K(G)$ and is defined as,
		\[K(G)=\sum\limits_{u,v \in V(G)} \dfrac{d(u,v)}{d(u)+d(v)}.\]
	\end{definition}
	
	\section*{Methods}
	
	 Inspired by the above distance-based indices, based on the actual bond lengths, here we introduce the following distanced based indices. In this new approach, we suggest that the actual bond lengths of three dimensional chemical structure should capture more information than two dimensional molecular graph structure of a particular drug. Also, Most of applications on topological indices, have used the hydrogen suppression of molecular graphs to get the results with relevant physical-chemical properties of the chemical structures. Also, they have considered the multi-bonds as single bond, in the molecule structure when they calculate the topological indices. But, we suggest that omitting some or all hydrogens, and considering multi-bonds as single bonds , the significant number of information of the molecule is lost. Computing the topological indices without omitting hydrogen and multi-bonds, would give more reliable results than computing indices of hydrogen suppressed structure.   
	 
	 In this study, we compare two computing methods of some topological indices of anti tuberculosis drugs. The methods are as listed below.
	 
	  \begin{enumerate}[label=\roman*.]
	 	\item \textbf{The traditional method}-
In this method, the following assumptions have been made.

\begin{itemize}
	\item The hydrogen suppressed molecule structure is studied.
	\item The multi-bonds are studied as the single bonds.
	\item The two dimensional molecular structure is studied.
	\item The graph theoretical distances of two dimensional(2D) chemical graphs are used.
\end{itemize}
	 	\item \textbf{The proposed method},- In this method 
	 	\begin{itemize}
	 		\item The hydrogen atoms aren't omitted.
	 		\item The multi bonds are studied as multi edges.
	 		\item Three dimensional real molecular structure is studied.
	 		\item The geometrical distances of the 3D chemical graphs are considered.
	 	\end{itemize}
	 \end{enumerate}
	 
	 In our work, the proposed method will be used for the comparison of existing degree based indices and novel distance based indices for anti tuberculosis drugs.
	 \section*{Results and discussion}
	 In this section, we compute the degree-based topological indices of the anti-tuberculosis drugs. Also, we calculate the correlation coefficient of those indices and physical properties of relevant drugs.  
	 
	 Now we define new distance based indices based on a property of an edge of the molecular graph. Let \(bl(u,v)\) be the weight of the edge \(uv\) of the molecular graph \(G\). In the context of molecular graph, we take \(bl(u,v)\) as the geometrical bond length of three dimensional molecular graph. We define the new distance based 3D indices, namely Enhanced Wiener index(EW), Enhanced Hyper Wiener index(EHW), Enhanced Second Weiner index(EWW), Enhanced Harary index(EHa), Enhanced Balaban index(EB), Enhanced Shultz index(ES), Enhanced Gutman index(EG) and Modified Klein index(K) as follows.
	\begin{definition}
		The Enhanced Wiener index is denoted by $EW(G)$ and is defined as,
		\[EW(G)=\sum\limits_{u,v \in V(G)} bl(u,v).\]
	\end{definition}

		\begin{definition}
		The Enhanced second hyper Wiener index is denoted by $EHW(G)$ and is defined as,
		\[EHW(G)=\sum\limits_{u,v \in V(G)} bl(u,v)^2.\]
	\end{definition}
	
	\begin{definition}
		The Enhanced Hyper Wiener index is denoted by $EHW(G)$ and is defined as,
		\[EWW(G)=\left( \sum\limits_{u,v \in V(G)} bl(u,v)+\sum\limits_{u,v \in V(G)} {bl(u,v)}^2\right) .\]

	\end{definition}
	
	\begin{definition}
		The Enhanced Harary index is denoted by $EHa(G)$ and is defined as,
		\[EHa(G)=\sum\limits_{u,v \in V(G)} \dfrac{1}{bl(u,v)}.\]
	\end{definition}
	
	\begin{definition}
		The Enhanced Schultz index is denoted by $ES(G)$ and is defined as,
		\[ES(G)=\sum\limits_{u,v \in V(G)} (d_G(u)+d_G(v))bl(u,v).\]
	\end{definition}
	
	\begin{definition}
		The Enhanced Gutman index is denoted by $EG(G)$ and is defined as,
		\[EG(G)=\sum\limits_{u,v \in V(G)} (d_G(u)\cdot d_G(v))bl(u,v).\]
	\end{definition}
	
	\begin{definition}
		The Enhanced  Balaban index is denoted by $M\!B(G)$ and is defined as,
		\[E\!B(G)=\sum\limits_{u,v \in V(G)} \dfrac{1}{bl(u,v)+1}.\]
	\end{definition}
	
	\begin{definition}
		The Enhanced Klein index is denoted by $K(G)$ and is defined as,
		\[EK(G)=\sum\limits_{u,v \in V(G)} \dfrac{bl(u,v)}{d(u)+d(v)}.\]
	\end{definition}
	
	\newpage
	\clearpage
	
	\begin{figure}[!ht]
		
		\begin{subfigure}{.33\textwidth}
			\centering
			\includegraphics[width=.95\linewidth]{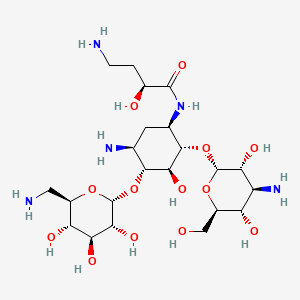}
			\caption{Amikacin \cite{Amikacin}}
			\label{fig:sfig11}
		\end{subfigure}%
		\begin{subfigure}{.33\textwidth}
			\centering
			\includegraphics[width=.95\linewidth]{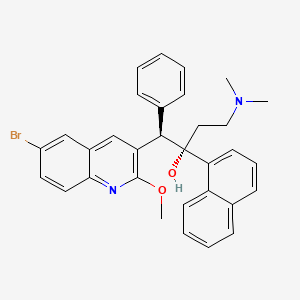}
			\caption{Bedaquiline \cite{Bedaquiline}}
			\label{fig:sfig22}
		\end{subfigure}%
		\begin{subfigure}{.33\textwidth}
			\centering
			\includegraphics[width=.95\linewidth]{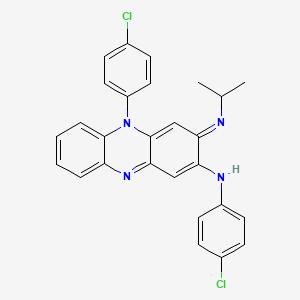}
			\caption{Clofazimine \cite{Clofazimine}}
			\label{fig:sfig33}
		\end{subfigure}%
		\vfill
		\begin{subfigure}{.33\textwidth}
			\centering
			\includegraphics[width=.95\linewidth]{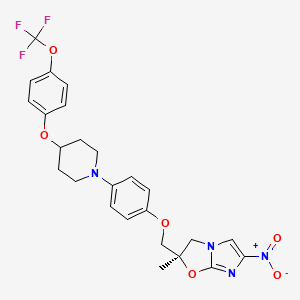}
			\caption{Delamanid \cite{Delamanid}}
			\label{fig:sfig44}
		\end{subfigure}
		\begin{subfigure}{.33\textwidth}
			\centering
			\includegraphics[width=.95\linewidth]{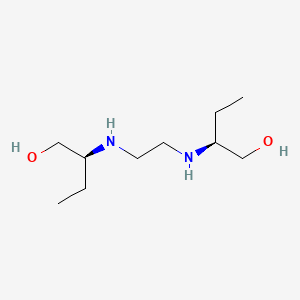}
			\caption{Ethambutol \cite{Ethambutol}}
			\label{fig:sfig55}
		\end{subfigure}%
		\begin{subfigure}{.33\textwidth}
			\centering
			\includegraphics[width=.95\linewidth]{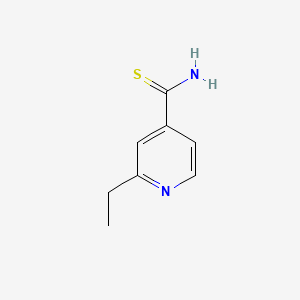}
			\caption{Ethionamide \cite{Ethionamide}}
			\label{fig:sfig66}
		\end{subfigure}
		\vfill
		\begin{subfigure}{.33\textwidth}
			\centering
			\includegraphics[width=.95\linewidth]{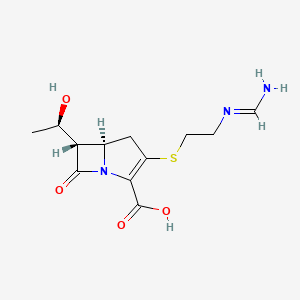}
			\caption{Imipenem \cite{Imipenem}}
			\label{fig:sfig77}
		\end{subfigure}%
		\begin{subfigure}{.33\textwidth}
			\centering
			\includegraphics[width=.95\linewidth]{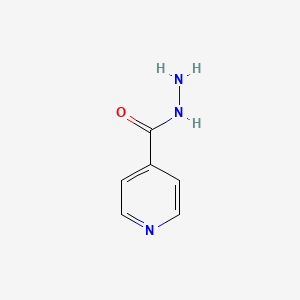}
			\caption{Isoniazid \cite{Isoniazid}}
			\label{fig:sfig88}
		\end{subfigure}%
		\begin{subfigure}{.33\textwidth}
			\centering
			\includegraphics[width=.95\linewidth]{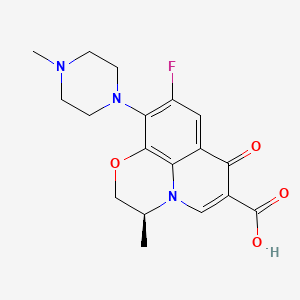}
			\caption{Levofloxacin \cite{Levofloxacin}}
			\label{fig:sfig99}
		\end{subfigure}\\
		\vfill
		\begin{subfigure}{.33\textwidth}
			\centering
			\includegraphics[width=.95\linewidth]{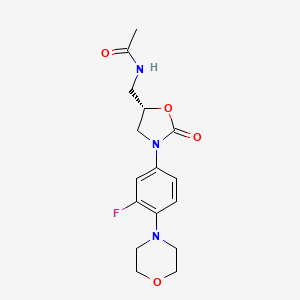}
			\caption{Linezolid \cite{Linezolid}}
			\label{fig:sfig10}
		\end{subfigure}%
		\begin{subfigure}{.33\textwidth}
			\centering
			\includegraphics[width=.95\linewidth]{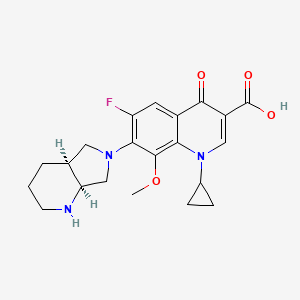}
			\caption{Moxifloxacin \cite{Moxifloxacin}}
			\label{fig:sfig111}
		\end{subfigure}%
		\begin{subfigure}{.33\textwidth}
			\centering
			\includegraphics[width=.95\linewidth]{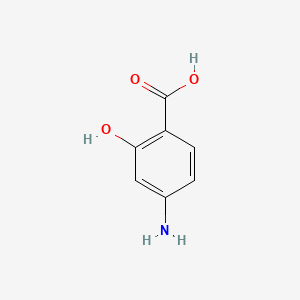}
			\caption{P-Aminosalicyclic acid \cite{p_Aminosalicylic}}
			\label{fig:sfig12}
		\end{subfigure}
		
		\caption{Chemical Structures of Medicines\\
			Resource:Chemical structures are taken from PubChem }
		\label{}
	\end{figure}
	
		\begin{figure}[ht!]
		
		\begin{subfigure}{.5\textwidth}
			\centering
			\includegraphics[width=1.5\linewidth]{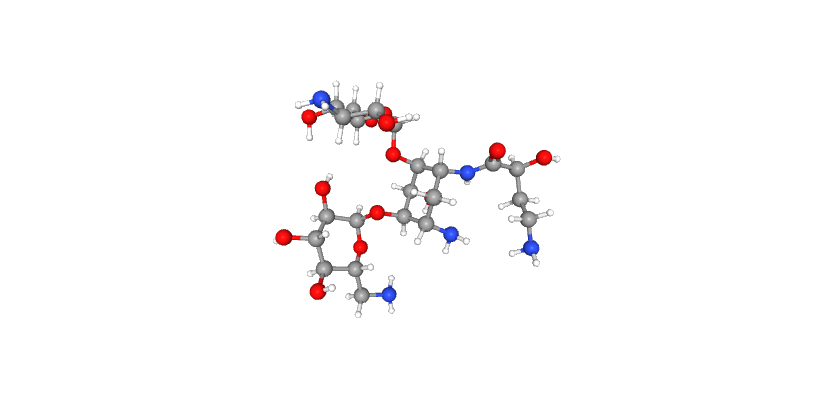}
			\caption{Amikacin}
			\label{fig:sfig1111}
		\end{subfigure}%
		\begin{subfigure}{.5\textwidth}
			\centering
			\includegraphics[width=.95\linewidth]{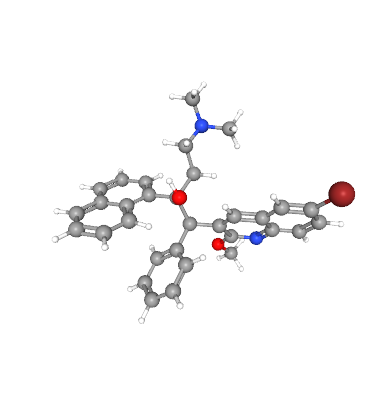}
			\caption{Bedaquiline}
			\label{fig:sfig222}
		\end{subfigure}\\
		\vfill
		\begin{subfigure}{.5\textwidth}
			\centering
			\includegraphics[width=.95\linewidth]{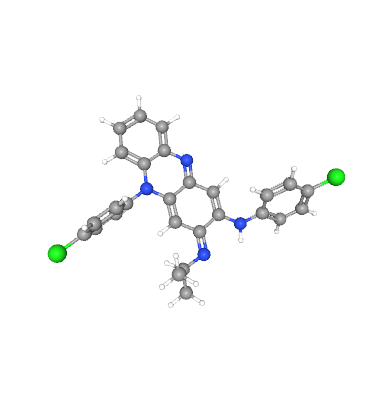}
			\caption{Clofazimine}
			\label{fig:sfig333}
		\end{subfigure}%
		\begin{subfigure}{.5\textwidth}
			\centering
			\includegraphics[width=.95\linewidth]{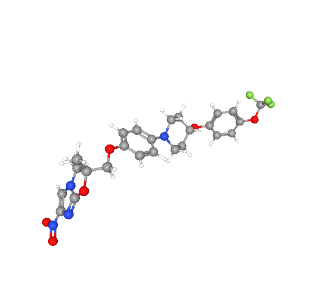}
			\caption{Delamanid}
			\label{fig:sfig444}
		\end{subfigure}%
	
\caption{Chemical Structures of Medicines\\
	Resource:Chemical structures are taken from PubChem }
\label{}

	\end{figure}
	
	\begin{figure}
		\ContinuedFloat
		\centering		
\begin{subfigure}{.5\textwidth}
	\centering
	\includegraphics[width=.95\linewidth]{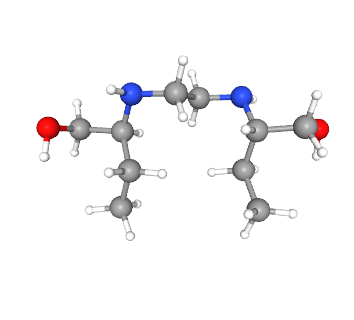}
	\caption{Ethambutol}
	\label{fig:sfig555}
\end{subfigure}%
\begin{subfigure}{.5\textwidth}
	\centering
	\includegraphics[width=.95\linewidth]{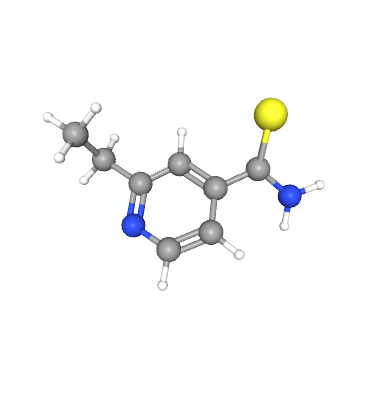}
	\caption{Ethionamide}
	\label{fig:sfig666}
\end{subfigure}\\
\vfill
\begin{subfigure}{.5\textwidth}
	\centering
	\includegraphics[width=.95\linewidth]{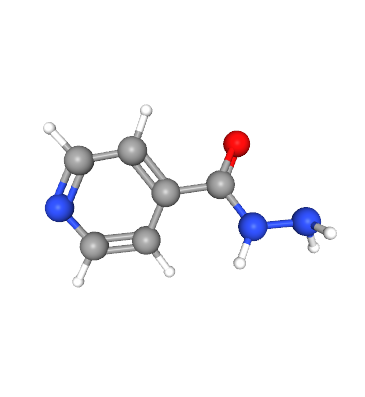}
	\caption{Isoniazid}
	\label{fig:sfig777}
\end{subfigure}%
\begin{subfigure}{.5\textwidth}
	\centering
	\includegraphics[width=.95\linewidth]{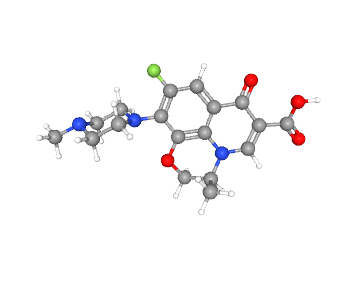}
	\caption{Levofloxacin}
	\label{fig:sfig999}
\end{subfigure}
\caption[]{Chemical Structures of Medicines (Ctd.)\\
	Resource:Chemical structures are taken from PubChem }
\label{}
	\end{figure}
	
	\begin{figure}\ContinuedFloat

\begin{subfigure}{.5\textwidth}
	\centering
	\includegraphics[width=.95\linewidth]{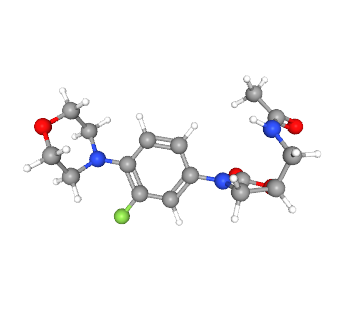}
	\caption{Linezolid}
	\label{fig:sfig101}
\end{subfigure}%
\begin{subfigure}{.5\textwidth}
	\centering
	\includegraphics[width=.95\linewidth]{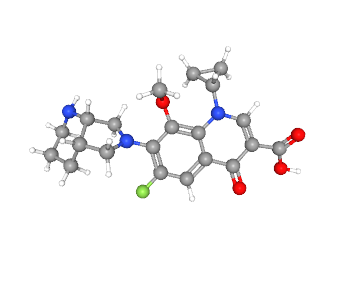}
	\caption{Moxifloxacin}
	\label{fig:sfig11111}
\end{subfigure}\\
\vfill
\begin{subfigure}{.5\textwidth}
	\centering
	\includegraphics[width=.95\linewidth]{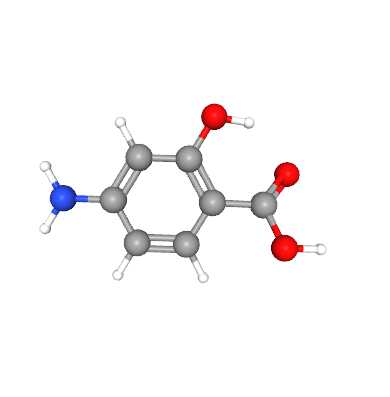}
	\caption{P-Aminosalicyclic}
	\label{fig:sfig1212}
\end{subfigure}%
\begin{subfigure}{.5\textwidth}
	\centering
	\includegraphics[width=.95\linewidth]{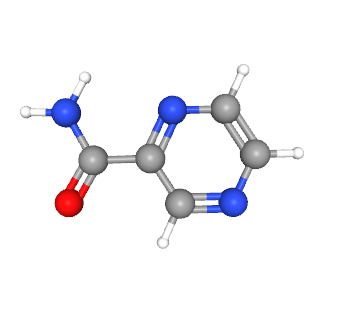}
	\caption{Pyrazinamide}
	\label{fig:sfig1313}
\end{subfigure}\\
\vfill
\begin{subfigure}{.5\textwidth}
	\centering
	\includegraphics[width=.95\linewidth]{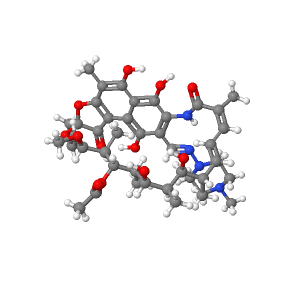}
	\caption{Rifampin(Cristal structure)}
	\label{fig:sfig1414}
\end{subfigure}%
\begin{subfigure}{.5\textwidth}
	\centering
	\includegraphics[width=.95\linewidth]{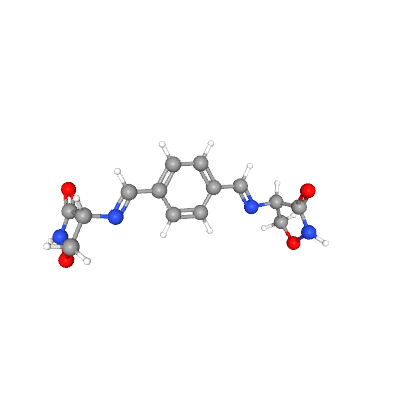}
	\caption{Terizidone}
	\label{fig:sfig1515}
\end{subfigure}
\caption{Chemical Structures of Medicines (Ctd.)\\
	Resource:Chemical structures are taken from PubChem }
\label{}
	\end{figure}
	
	
	We investigate the relationships between existing degree-based indices, existing distance/degree distance indices, and new distance/degree distance indices with physical properties of antituberculosis drugs, including Boiling point (BP), Melting point (MP), Flash point (FP), Enthalpy of vaporization (EV), Molar refraction (MR), Polarizability (PL), Surface tension(ST) and Molar volume (MV).
	
	\textbf{Regression models for the novel distance based indices:}\\
	
	Linear regression entails fitting a linear equation to observable data to model the connection between two variables. This theory posits that the first variable serves as an explanatory variable, whereas the second variable functions as a dependent variable. The regression model enables the calculation of linear regression values for each topological index, as demonstrated below:
	
	\[P=a+b\left[ T\!I\right] \]
	
	The dependent variable is represented by \(P\), while the independent parameter is represented by \(T\!I\).
	The parameter \(P\) will denote the physical or chemical properties of the pharmaceutical employed in the treatment of tuberculosis, while the parameter \(T\!I\) will signify the estimated values of characteristics or topological indices.
	\begin{multicols}{2}

		\begin{enumerate}[label=\arabic*.]
			\item Boiling Point=$70.4+6.963\left[ EW(G)\right] $\\
			Flash point =$43.5+3.738\left[ EW(G)\right] $\\
			Enthalphy of vaporization=$28.5+0.908\left[EW(G) \right]$\\
			Molar refraction=$ -6.3+1.526\left[ EW(G)\right] $\\
			Polarizability=$-2.51+0.6053\left[ EW(G)\right]  $\\
				Molar Volume=$-6.4+4.133\left[ EW(G)\right]  $\\
			
			\item Boiling Point=$90.1+5.077\left[ EHW(G)\right] $\\
			Flash point =$55.0+2.716\left[ EHW(G) \right] $\\
			Enthalphy of vaporization=$29.2+0.679\left[EHW(G) \right]$\\
			Molar refraction=$ -3.4+1.129\left[ EHW(G)\right]$\\
			Polarizability=$-1.35+0.4476\left[ EHW(G)\right]  $\\
			Molar Volume=$1.1+3.062\left[ EHW(G)\right]  $\\
			
			\item Boiling Point=$74.7+2.971\left[ EWW(G)\right] $\\
			Flash point =$44.6+1.603\left[ EWW(G)\right] $\\
			Enthalphy of vaporization=$28.9+0.3888\left[EWW(G) \right]$\\
			Molar refraction=$ -6.1+0.6566\left[ EWW(G)\right] $\\
			Polarizability=$-2.46+0.2604\left[ EWW(G)\right]  $\\
				Molar Volume=$-6.3+1.780\left[ EWW(G)\right]  $\\
			
			\item Boiling Point=$147+8.91\left[ EHa(G)\right] $\\
            Flash point =$93.3+4.59\left[ EHa(G)\right] $\\
			Enthalphy of vaporization=$44.0+1.053\left[EHa(G) \right]$\\
			Molar refraction=$ 36.7+1.425\left[ EHa(G)\right] $\\
			Polarizability=$14.54+0.565\left[ EHa(G)\right]  $\\
			Molar Volume=$102.2+4.05\left[ EHa(G)\right]  $\\
			
			\item Boiling Point=$64.6+20.90\left[ EB(G)\right] $\\
			Flash point =$44.9+11.02\left[ EB(G) \right] $\\
			Enthalphy of vaporization=$28.4+2.702\left[EB(G) \right]$\\
			Molar refraction=$-4.1+4.424\left[EB(G)\right] $\\
			Polarizability=$-1.63+1.754\left[ EB(G)\right]  $\\
			Molar Volume=$0.5+11.94\left[ EB(G)\right]  $\\
			
			\item Boiling Point=$351+0.483\left[ ES(G)\right] $\\
			Flash point =$213.0+0.211\left[ ES(G)\right] $\\
			Enthalphy of vaporization=$76.7+0.0398\left[ES(G) \right]$\\
			Molar refraction=$ 61.2+0.0875\left[ ES(G)\right] $\\
			Polarizability=$24.3+0.0347\left[ ES(G)\right]  $\\
			Molar Volume=$185.9+0.212\left[ ES(G)\right]  $\\

			\item Boiling Point=$135.5+0.579\left[ EG(G)\right] $\\
			Enthalphy of vaporization=$41.6+0.07\left[EG(G) \right]$\\
			Flash point =$77.4+0.3125\left[ EG(G)\right] $\\
			Molar refraction=$ -0.6+0.1387\left[ EG(G)\right] $\\
			Polarizability=$-0.25+0.0550\left[ EG(G)\right]  $\\
			Molar Volume=$12.2+0.3712\left[ EG(G)\right]  $\\
		\end{enumerate}

	\end{multicols}

	These distance/ degree-distance indices do not show linear relationships with \(MP\) and \(S\!T\).

	\newpage
\clearpage
	Table \ref{table1} shows the computed degree based topological indices for the antituberculosis drugs using the proposed method.
	\begin{table}[!htb]
		\centering
		\caption{Values of degree based indices for the proposed method}
		\label{table1}
		\begin{tabular}{|c|c|c|c|c|c|c|c|c|c|c|}
			\hline
			\textbf{Name of Medicine} & \textbf{M1} & \textbf{M2} & \textbf{H} & \textbf{ABC} & \textbf{R} & \textbf{S} & \textbf{GA} & \textbf{HM} & \textbf{ZG3} & \textbf{F } \\ \hline
			\textbf{Amikacin} & 482 & 659 & 31.898 & 115.568 & 35.699 & 36.185 & 76.446 & 2974 & 140 & 1656  \\ \hline
			\textbf{Bedaquiline} & 563 & 914 & 26.495 & 149.257 & 29.711 & 33.112 & 77.52 & 3953 & 105 & 2125  \\ \hline
			\textbf{Clofazizime} & 493 & 823 & 22.057 & 133.181 & 24.464 & 28 & 67.163 & 3513 & 81 & 1867  \\ \hline
			\textbf{Delamanid} & 491 & 754 & 24.56 & 126.377 & 27.479 & 30.386 & 69.969 & 3305 & 111 & 1797  \\ \hline
			\textbf{Ethambutol} & 210 & 266 & 13.893 & 48.089 & 16.181 & 15.902 & 32.462 & 1266 & 74 & 734  \\ \hline
			\textbf{Ethionamide} & 160 & 246 & 8.26 & 41.181 & 9.267 & 10.088 & 22.977 & 1076 & 36 & 584  \\ \hline
			\textbf{Isoniazid} & 134 & 210 & 6.99 & 35.634 & 7.677 & 8.5 & 19.643 & 900 & 26 & 480  \\ \hline
			\textbf{Levofloxacin} & 352 & 534 & 18.071 & 89.446 & 20.28 & 22.28 & 50.45 & 2354 & 84 & 1286  \\ \hline
			\textbf{Linezolid} & 315 & 479 & 15.919 & 80.122 & 17.862 & 19.641 & 44.984 & 2109 & 75 & 1151  \\ \hline
			\textbf{Moxifloxacin} & 396 & 601 & 20.486 & 100.722 & 23.069 & 25.023 & 56.726 & 2652 & 94 & 1450  \\ \hline
			\textbf{p\_Aminosalicylic} & 140 & 226 & 7.652 & 37.734 & 8.272 & 9.043 & 20.779 & 958 & 24 & 506  \\ \hline
			\textbf{Pyrazinamide} & 124 & 198 & 6.086 & 32.898 & 6.525 & 7.557 & 18.078 & 838 & 24 & 442  \\ \hline
			\textbf{Rifampin} & 454 & 730 & 23.319 & 48.116 & 24.319 & 28.586 & 68.849 & 3028 & 58 & 1568  \\ \hline
			\textbf{Terizidone} & 274 & 430 & 13.502 & 71.33 & 14.981 & 16.734 & 38.964 & 1864 & 58 & 1004  \\ \hline
		\end{tabular}
	\end{table}
	
	According to Table \ref{table1},  the values of degree based indices are in the range 6.086-3953 for the proposed method.

	The Table \ref{table2} shows the physical properties for anti-tuberculosis drugs. Those physical properties are used to compute correlations with topological indices.
	\begin{table}[!ht]
		\caption{Physical properties of Anti-tuberculosis drugs}
		\label{table2}
		\centering
		\begin{tabular}{|c|c|c|c|c|c|c|c|c|}
			\hline
			\textbf{Name of Medicine} & \textbf{BP} & \textbf{MP} & \textbf{FP} & \textbf{EV} & \textbf{MR} & \textbf{PL} & \textbf{ST} & \textbf{MV } \\ \hline
			\textbf{Amikacin} & 981.8 & 203.5 & 547.6 & 162.2 & 134.9 & 53.5 & 103.3 & 363.9  \\ \hline
			\textbf{Bedaquiline} & 702.7 & 176 & 378.8 & 108 & 156.2 & 61.9 & 52.6 & 420.1  \\ \hline
			\textbf{Clofazimine} & 566.9 & 210 & 296.7 & 85.1 & 136.2 & 54 & 47.1 & 366.1  \\ \hline
			\textbf{Delanamide} & 653.7 & 193 & 349.1 & 96.3 & 127.7 & 50.6 & 50 & 368  \\ \hline
			\textbf{Ethambutol} & 345.3 & 89 & 113.7 & 68.3 & 58.6 & 23.2 & 38.1 & 207  \\ \hline
			\textbf{Ethionamide} & 247.9 & 163 & 103.7 & 46.5 & 49 & 19.4 & 39.8 & 142  \\ \hline
			\textbf{Isoniazid} & 251.97 & 172 & 251 & NA & 36.9 & 14.6 & 57.8 & 110.2  \\ \hline
			\textbf{Levofloxacin} & 571.5 & 224 & 299.4 & 90.1 & 91.1 & 36.1 & 70.3 & 244  \\ \hline
			\textbf{Linezolid} & 585.5 & 177 & 307.9 & 87.5 & 83 & 32.9 & 47.7 & 259  \\ \hline
			\textbf{Moxifloxacin} & 636 & 270 & 338.7 & 98.8 & 101.8 & 40.4 & 60.6 & 285  \\ \hline
			\textbf{P-Aminosalicylic acid} & 380.8 & 145 & 184.1 & 66.3 & 39.3 & 15.6 & 83.4 & 102.7  \\ \hline
			\textbf{Pyrazinamide} & 173.3 & 190 & 119.1 & 54.1 & 31.9 & 12.6 & 60.7 & 87.7  \\ \hline
			\textbf{Rifampin} & 937.4 & 183 & 561.3 & 153.5 & 213.1 & 84.5 & 48 & 611.7  \\ \hline
			\textbf{Terizidone} & NA & 175 & NA & NA & 76.1 & 30.2 & 62.5 & 198.9  \\ \hline
		\end{tabular}
		\text{Resource: These values are taken from Chemspider.}
	\end{table}
	
		\newpage
	\clearpage
	
	Also,	Table \ref{table5} shows the correlation coefficients of degree-based indices and physical properties for the proposed method.
	\begin{table}[!ht]
		\centering
		\caption{Correlation Coefficients of Degree based indices and physical properties}
		\label{table5}
		\begin{tabular}{|l|l|l|l|l|l|l|l|l|}
			\hline
			\textbf{} & \textbf{BP} & \textbf{MP} & \textbf{FP} & \textbf{EV} & \textbf{MR} & \textbf{PL} & \textbf{ST} & \textbf{MV} \\ \hline
			\textbf{M1} & 0.855 & 0.454 & 0.782 & 0.741 & 0.894 & 0.894 & 0.052 & 0.869 \\ \hline
			\textbf{M2} & 0.808 & 0.464 & 0.748 & 0.677 & 0.898 & 0.898 & -0.0014 & 0.868 \\ \hline
			\textbf{H} & 0.922 & 0.394 & 0.839 & 0.86 & 0.856 & 0.856 & 0.225 & 0.838 \\ \hline
			\textbf{ABC} & 0.609 & 0.467 & 0.512 & 0.429 & 0.602 & 0.602 & 0.097 & 0.557 \\ \hline
			\textbf{R} & 0.904 & 0.387 & 0.813 & 0.835 & 0.829 & 0.829 & 0.226 & 0.811 \\ \hline
			\textbf{S} & 0.912 & 0.418 & 0.83 & 0.834 & 0.877 & 0.877 & 0.171 & 0.857 \\ \hline
			\textbf{GA} & 0.903 & 0.439 & 0.832 & 0.815 & 0.909 & 0.909 & 0.104 & 0.887 \\ \hline
			\textbf{HM} & 0.813 & 0.462 & 0.745 & 0.682 & 0.885 & 0.885 & 0.003 & 0.856 \\ \hline
			\textbf{ZG3} & 0.768 & 0.365 & 0.642 & 0.668 & 0.615 & 0.615 & 0.251 & 0.603 \\ \hline
			\textbf{F } & 0.815  & 0.459  & 0.74  & 0.683  & 0.872  & 0.872  & 0.018  & 0.843  \\ \hline
		\end{tabular}
	\end{table}
	
	Table \ref{table3} gives the computed values of the existing distance /degree distance based indices for the selected anti-tuberculosis drugs.In this process, the edge lengths are considered as the graph theoretical distances. 
		\begin{table}[!ht]
		\centering
		\caption{Distance based indices for anti-tuberculosis drugs when distances are graph theoretic distances}
		\label{table3}
		\begin{tabular}{|l|l|l|l|l|l|l|}
			\hline
			\textbf{Name of Medicine } & \textbf{W} & \textbf{HW} & \textbf{WW} & \textbf{BB} & \textbf{Sch} & \textbf{Gut} \\ \hline
			\textbf{Amikacin } & 84 & 84 & 168 & 84 & 482 & 659 \\ \hline
			\textbf{Bedaquiline } & 84 & 84 & 168 & 84 & 563 & 914 \\ \hline
			\textbf{Clofazizime } & 73 & 73 & 146 & 73 & 493 & 823 \\ \hline
			\textbf{Delamanid } & 76 & 76 & 152 & 76 & 491 & 754 \\ \hline
			\textbf{Ethambutol } & 37 & 37 & 74 & 37 & 210 & 266 \\ \hline
			\textbf{Ethionamide } & 25 & 25 & 50 & 25 & 160 & 246 \\ \hline
			\textbf{Isoniazid } & 21 & 21 & 42 & 21 & 134 & 210 \\ \hline
			\textbf{Levofloxacin } & 55 & 55 & 110 & 55 & 352 & 534 \\ \hline
			\textbf{Linezolid } & 49 & 49 & 98 & 49 & 315 & 479 \\ \hline
			\textbf{Moxifloxacin } & 62 & 62 & 124 & 62 & 396 & 601 \\ \hline
			\textbf{p\_Aminosalicylic } & 22 & 22 & 44 & 22 & 140 & 226 \\ \hline
			\textbf{Pyrazinamide } & 19 & 19 & 38 & 19 & 124 & 198 \\ \hline
			\textbf{Rifampin } & 71 & 71 & 142 & 71 & 454 & 730 \\ \hline
			\textbf{Terizidone } & 42 & 42 & 84 & 42 & 274 & 430 \\ \hline
		\end{tabular}
	\end{table}

		As in Table \ref{table3}, when we consider the graph theoretic distance Weiner index, Hyper Weiner index, second hyper Weiner index, Schultz index and Gutman index,  \(d(u,v)=1\). Then, Weiner index, Hyper Weiner index gives the same values. Also, Shultz index and Gutman Index automatically become the first and second Zagreb indices respectively. In the context of chemical graphs, the Schultz index and Gutman index is meaningless since they coincide with the first and second Zagreb indices when \(d(u,v)\) equals graph theoretical distance.\\
	
		\pagebreak
	\clearpage
	\newpage
	Table \ref{table4} presents the computed values of the novel distance/degree-distance based topological indices of the anti-tuberculosis drugs for the proposed drugs.
	
	\begin{table}[!ht]
		\centering
		
		\caption{Novel distance/degree distance based topological indices values of medicines}
		\label{table4}
		\begin{tabular}{|c|c|c|c|c|c|c|c|}
			\hline
			\textbf{Name of Medicine} & \textbf{EW} & \textbf{EHW} & \textbf{EWW} & \textbf{EHa} & \textbf{EB} & \textbf{ES} & \textbf{EG } \\ \hline
			\textbf{Amikacin} & 106.826 & 139.916 & 246.742 & 68.096 & 37.326 & 640.1 & 925.689  \\ \hline
			\textbf{Bedaquiline} & 108.992 & 144.008 & 253 & 65.957 & 36.776 & 747.862 & 1253.408  \\ \hline
			\textbf{Clofazizime} & 94.553 & 126.143 & 220.696 & 55.75 & 31.289 & 659.764 & 1129.253  \\ \hline
			\textbf{Delamanid} & 98.8 & 130.276 & 229.076 & 59.35 & 33.199 & 650.598 & 1029.164  \\ \hline
			\textbf{Ethambutol} & 45.285 & 56.925 & 102.21 & 30.999 & 16.768 & 267.159 & 361.86  \\ \hline
			\textbf{Ethionamide} & 32.396 & 42.974 & 75.37 & 19.755 & 10.971 & 212.636 & 338.29  \\ \hline
			\textbf{Isoniazid} & 26.501 & 26.501 & 60.439 & 16.909 & 9.329 & 173.582 & 280.853  \\ \hline
			\textbf{Levofloxacin} & 71.011 & 93.224 & 164.235 & 43.35 & 24.137 & 465.607 & 732.138  \\ \hline
			\textbf{Linezolid} & 62.989 & 82.35 & 145.339 & 82.35 & 21.557 & 413.231 & 648.148  \\ \hline
			\textbf{Moxifloxacin} & 79.752 & 104.49 & 184.242 & 49.132 & 27.28 & 523.354 & 826.327  \\ \hline
			\textbf{p\_Aminosalicylic} & 27.849 & 35.875 & 63.724 & 17.724 & 9.766 & 183.369 & 305.664  \\ \hline
			\textbf{Pyrazinamide} & 24.279 & 31.404 & 55.683 & 15.072 & 8.375 & 161.539 & 263.782  \\ \hline
			\textbf{Rifampin} & 99.659 & 140.702 & 240.361 & 50.882 & 29.598 & 18.653 & 1008.919  \\ \hline
			\textbf{Terizidone} & 53.972 & 70.432 & 124.403 & 1.649 & 18.473 & 360.422 & 584.208  \\ \hline
		\end{tabular}
	\end{table}
	In the Table \ref{table4}, the computed novel distance/ degree distance indices were computed using the proposed method. The lengths of edges were taken as the actual bond lengths of three dimensional chemical structure of the relevant anti-tuberculosis drugs. 

The correlation values of novel distance/ degree distance indices with physical properties of the given anti-tuberculosis drugs are presented in the Table \ref{table6}.
	\begin{table}[!ht]
		\centering
		\caption{Correlation of new distance/degree-distance based indices and physical properties}
		\label{table6}
		\begin{tabular}{|c|c|c|c|c|c|c|c|c|}
			\hline
			\textbf{} & \textbf{BP} & \textbf{MP} & \textbf{FP} & \textbf{EV} & \textbf{MR} & \textbf{PL} & \textbf{ST} & \textbf{MV} \\ \hline
			\textbf{EW} & 0.905 & 0.437 & 0.835 & 0.818 & 0.918 & 0.918 & 0.099 & 0.898 \\ \hline
			\textbf{EHW} & 0.912 & 0.431 & 0.838 & 0.830 & 0.937 & 0.937 & 0.083 & 0.918 \\ \hline
			\textbf{EWW} & 0.910 & 0.436 & 0.843 & 0.825 & 0.931 & 0.931 & 0.090 & 0.911 \\ \hline
			\textbf{EHa} & 0.785 & 0.324 & 0.695 & 0.645 & 0.654 & 0.654 & 0.011 & 0.671 \\ \hline
			\textbf{EB} & 0.888 & 0.430 & 0.805 & 0.795 & 0.869 & 0.869 & 0.139 & 0.847 \\ \hline
			\textbf{ES} & 0.454 & 0.421 & 0.341 & 0.269 & 0.378 & 0.378 & 0.169 & 0.331 \\ \hline
			\textbf{EG} & 0.816 & 0.466 & 0.756 & 0.690 & 0.902 & 0.902 & -0.002 & 0.871 \\ \hline
		\end{tabular}
	\end{table}
	\clearpage 
Table \ref{table6} presents correlations values of the physical properties of the anti-tuberculosis drugs with the novel distance/ degree-distance based indices.
		\begin{figure}[ht!]
		\caption{Correlation of physical properties and novel distance/ degree-distance based topological indices. }
		\begin{subfigure}{.5\textwidth}
			\centering
			\includegraphics[width=.8\linewidth]{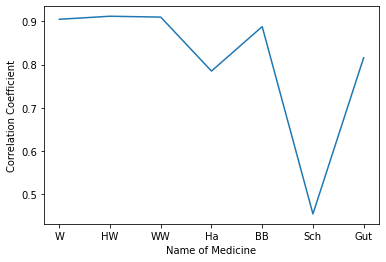}
			\subcaption{Boiling point}
			\label{fig:sfig1}
		\end{subfigure}%
		\begin{subfigure}{.5\textwidth}
			\centering
			\includegraphics[width=.8\linewidth]{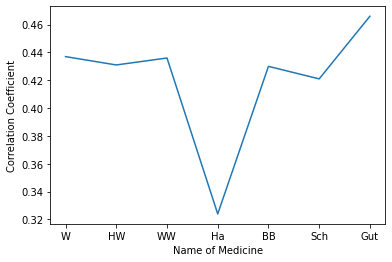}
			\caption{Melting point}
			\label{fig:sfig2}
		\end{subfigure}
		
		\vfill
		\begin{subfigure}{.5\textwidth}
			\centering
			\includegraphics[width=.8\linewidth]{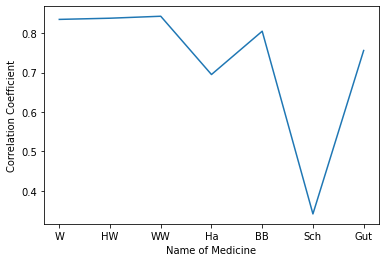}
			\caption{Flash point}
			\label{fig:sfig3}
		\end{subfigure}%
		\begin{subfigure}{.5\textwidth}
			\centering
			\includegraphics[width=.8\linewidth]{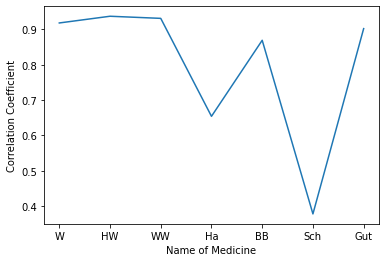}
			\caption{Enthalpy of Evaporation }
			\label{fig:sfig4}
		\end{subfigure}%
		
		\vfill
		\begin{subfigure}{.5\textwidth}
			\centering
			\includegraphics[width=.8\linewidth]{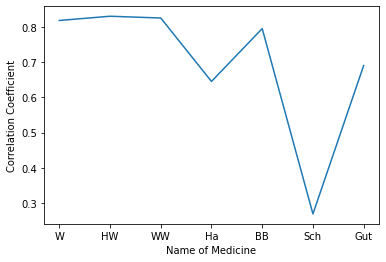}
			\caption{Molar refraction}
			\label{fig:sfig5}
		\end{subfigure}%
		\begin{subfigure}{.5\textwidth}
			\centering
			\includegraphics[width=.8\linewidth]{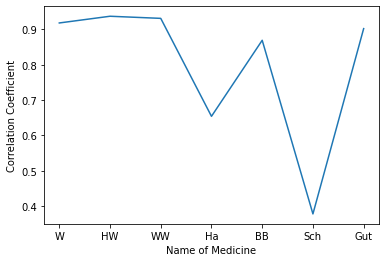}
			\caption{Polarizability}
			\label{fig:sfig6}
		\end{subfigure}
		\vfill
		\begin{subfigure}{.5\textwidth}
			\centering
			\includegraphics[width=.8\linewidth]{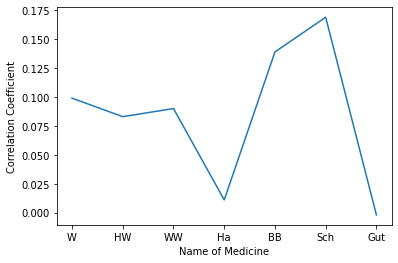}
			\caption{Surface tension}
			\label{fig:sfig7}
		\end{subfigure}%
		\begin{subfigure}{.5\textwidth}
			\centering
			\includegraphics[width=.8\linewidth]{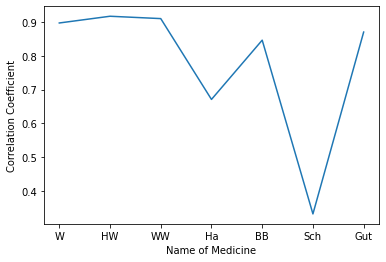}
			\caption{Molar volume}
			\label{fig:sfig8}
		\end{subfigure}%
		
	\end{figure}
	
\begin{figure}[!ht]
		
		\centering
		\includegraphics[width=0.7\textwidth]{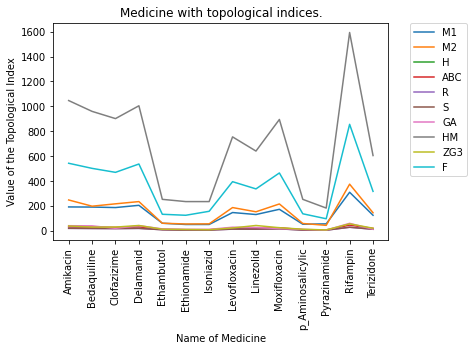}
		\caption{Comparison of degree based topological indices for the traditional method }
		\label{fig:fig4}
	\end{figure}

	Figure \ \ref{fig:fig4} \  illustrates values of the chosen degree based topological indices for each anti-tuberculosis drug for traditional method. 
	
\begin{figure}[!ht]
		
		\centering
		\includegraphics[width=0.7\textwidth]{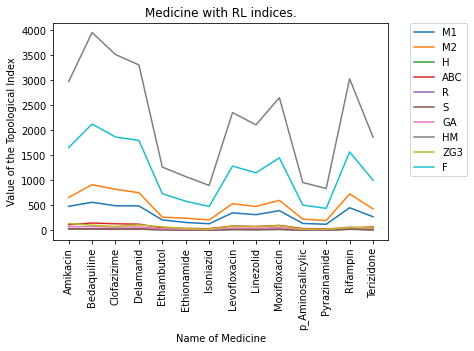}
		\caption{Comparison of degree based topological indices for the proposed method }
		\label{fig:fig5}
	\end{figure}

	According to Figure \ \ref{fig:fig4}, the values of the topological indices vary within \(0-1600\) corresponding to the given drugs. Figure \ \ref{fig:fig5} shows values of degree based topological indices for anti-tuberculosis drugs for the proposed method.
	
	According to Figure \ \ref{fig:fig5}, the values of degree based indices for the proposed method are within the range \(0-4000\) for relevant drugs.
	
\begin{figure}[htb]
		
		\centering
		\includegraphics[width=0.8\textwidth]{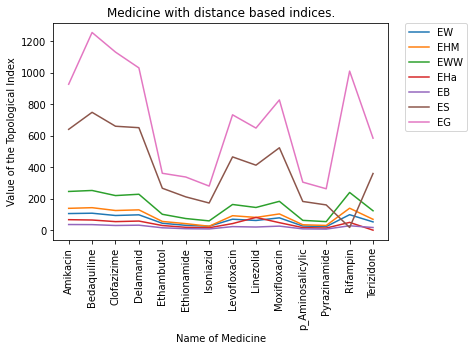}
		\caption{Comparison of distance based topological indices for the proposed method }
		\label{fig:fig6}
	\end{figure}

	Figure \ \ref{fig:fig6} \ illustrates the values of the novel distance/ degree-distance topological indices for the proposed method for the given anti-tuberculosis medicines.
	
According to Figure \ref{fig:fig6}, \ we can see the values of novel distance /degree-distance topological indices have been varied within \(0-1200\).

\subsubsection*{Computation of statistical parameters.}		

Table \ref{table7} - \ref{table13} \ show the statistical parameters for the linear QSPR model for \(EW(G),\ EHW(G),\ EWW(G),\ EHa(G),\ EB(G)\), \( ES(G)\) and \(EG(G)\). In these tables, \(N-\) number of observations, \(a,b-\) regression coefficients, \(r-\) correlation coefficient, \(R^2(adj)-\) coefficient of determination(adjusted).
	\begin{table}[!ht]
		\centering
		\caption{Statistical parameters for the  linear QSPR model for $EW(G)$}
		\label{table7}
		\begin{tabular}{|l|l|l|l|l|l|l|l|l|}
			\hline
			\textbf{} & \textbf{$N$} & \textbf{$a$} & \textbf{$b$} & \textbf{$r$} & \textbf{$R^2(adj)$} & \textbf{$F$ value} & \textbf{$p$ value } & \textbf{Indicator} \\ \hline
			\textbf{BP} & 13 & 70.4 & 6.963 & 0.905 & 0.8033 & 50 & 0& Significant\\ \hline
			\textbf{MP} & 14 & 146.5 & 0.557 & 0.437 & 0.1231 & 2.82 & 0.119 & Not Significant\\ \hline
			\textbf{FP} & 13 & 43.5 & 3.738 & 0.835 & 0.6696 & 25.32 & 0 & Significant\\ \hline
			\textbf{EV} & 12 & 28.5 & 0.908 & 0.818 & 0.6362 & 20.24 & 0.001 & Significant \\ \hline
			\textbf{MR} & 14 & -6.3 & 1.526 & 0.918 & 0.8300& 64.48 & 0  &Significant\\ \hline
			\textbf{PL} & 14 & -2.51 & 0.6053 & 0.918 & 0.8300 & 64.49 & 0 &Significant \\ \hline
			\textbf{ST} & 14 & 55.1 & 0.055 & 0.099 & 0 & 0.1200 & 0.735 &Not Significant \\ \hline
			\textbf{MV} & 14 & -6.4 & 4.133 & 0.898 & 0.7899 & 49.86 & 0  &Significant\\ \hline
		\end{tabular}
	\end{table}

	\begin{table}[!ht]
		\centering
		\caption{Statistical parameters for the  linear QSPR model for $EHW(G)$}
		\begin{tabular}{|l|l|l|l|l|l|l|l|l|}
			\hline
			\textbf{} & \textbf{$N$} & \textbf{$a$} & \textbf{$b$} & \textbf{$r$} & \textbf{$R^2(ad\!j)$} & \textbf{$F$- value} & \textbf{$p$- value } & \textbf{Indicator} \\ \hline
			\textbf{BP} & 13 & 90.1 & 5.077 & 0.912 & 0.8157 & 54.1 & 0 & Significant  \\ \hline
			\textbf{MP} & 14 & 148.7 & 0.398 & 0.431 & 0.1182 & 2.74 & 0.124 & Not significant  \\ \hline
			\textbf{FP} & 13 & 55 & 2.716 & 0.838 & 0.6746 & 25.88 & 0 & Significant  \\ \hline
			\textbf{EV} & 12 & 29.2 & 0.679 & 0.83 & 0.6575 & 22.12 & 0.001 & Significant  \\ \hline
			\textbf{MR} & 14 & -3.4 & 1.129 & 0.937 & 0.8684 & 86.75 & 0 & Significant  \\ \hline
			\textbf{PL} & 14 & -1.35 & 0.4476 & 0.937 & 0.8684 & 86.78 & 0 & Significant  \\ \hline
			\textbf{ST} & 14 & 55.8 & 0.033 & 0.083 & 0 & 0.08 & 0.779 & Not significant  \\ \hline
			\textbf{MV} & 14 & 1.1 & 3.062 & 0.918 & 0.8296 & 64.28 & 0 & Significant  \\ \hline
		\end{tabular}
	\end{table}

	\begin{table}[!ht]
		\centering
		\caption{Statistical parameters for the  linear QSPR model for $EWW(G)$}
		\label{table9}
		\begin{tabular}{|l|l|l|l|l|l|l|l|l|}
			\hline
			\textbf{} & \textbf{$N$} & \textbf{$a$} & \textbf{$b$} & \textbf{$r$} & \textbf{$R^2(ad\!j)$} & \textbf{$F$-value} & \textbf{$p$-value} & \textbf{Indicator} \\ \hline
			\textbf{BP} & 13 & 74.7 & 2.971 & 0.91 & 0.8119 & 52.81 & 0 & Significant  \\ \hline
			\textbf{MP} & 14 & 147.1 & 0.236 & 0.436 & 12.25 & 2.81 & 0.119 & Not significant  \\ \hline
			\textbf{FP} & 13 & 44.6 & 1.603 & 0.843 & 0.6846 & 27.05 & 0 & Significant  \\ \hline
			\textbf{EV} & 12 & 28.9 & 0.389 & 0.825 & 0.6492 & 21.35 & 0.001 & Significant  \\ \hline
			\textbf{MR} & 14 & -6.1 & 0.6566 & 0.931 & 0.8548 & 77.54 & 0 & Significant  \\ \hline
			\textbf{PL} & 14 & -2.46 & 0.2604 & 0.931 & 0.8548 & 77.55 & 0 & Significant  \\ \hline
			\textbf{ST} & 14 & 55.4 & 0.0211 & 0.09 & 0 & 0.1 & 0.759 & Not significant  \\ \hline
			\textbf{MV} & 14 & -6.3 & 1.78 & 0.911 & 0.8153 & 58.39 & 0 & Significant  \\ \hline
		\end{tabular}
	\end{table}

\begin{table}[!ht]
	\centering
	\caption{Statistical parameters for the  linear QSPR model for $EHa(G)$}
	\label{table10}
	\begin{tabular}{|l|l|l|l|l|l|l|l|l|}
		\hline
		\textbf{} & \textbf{$N$} & \textbf{$a$} & \textbf{$b$} & \textbf{$r$} & \textbf{$R^2(ad\!j)$} & \textbf{$F$-value} & \textbf{$p$-value } & \textbf{Indicator} \\ \hline
		\textbf{BP} & 13 & 147 & 8.91 & 0.785 & 0.5822 & 17.72 & 0.001 & Significant  \\ \hline
		\textbf{MP} & 14 & 161.3 & 0.542 & 0.324 & 0.0304 & 1.41 & 0.258 & Not significant  \\ \hline
		\textbf{FP} & 13 & 93.3 & 4.59 & 0.695 & 0.4357 & 10.27 & 0.008 & Significant  \\ \hline
		\textbf{EV} & 12 & 44 & 1.053 & 0.645 & 0.3583 & 7.14 & 0.023 & Significant  \\ \hline
		\textbf{MR} & 14 & 36.7 & 1.425 & 0.654 & 0.3796 & 8.95 & 0.011 & Significant  \\ \hline
		\textbf{PL} & 14 & 14.54 & 0.565 & 0.654 & 0.3794 & 8.95 & 0.011 & Significant  \\ \hline
		\textbf{ST} & 14 & 58.38 & 0.008 & 0.011 & 0 & 0 & 0.971 & Not significant  \\ \hline
		\textbf{MV} & 14 & 102.2 & 4.05 & 0.671 & 0.4038 & 9.8 & 0.009 & Significant  \\ \hline
	\end{tabular}
	\end{table}
	
\begin{table}[!ht]
	\centering
	\caption{Statistical parameters for the  linear QSPR model for $EB(G)$}
	\label{table11}
	\begin{tabular}{|l|l|l|l|l|l|l|l|l|}
		\hline
		\textbf{} & \textbf{$N$} & \textbf{$a$} & \textbf{$b$} & \textbf{$r$} & \textbf{$R^2(ad\!j)$} & \textbf{$F$-value} & \textbf{$p$-value } & \textbf{Indicator} \\ \hline
		\textbf{BP} & 13 & 64.6 & 20.9 & 0.888 & 0.769 & 40.94 & 0 & Significant  \\ \hline
		\textbf{MP} & 14 & 145.8 & 1.68 & 0.43 & 0.1173 & 2.73 & 0.125 & Not significant  \\ \hline
		\textbf{FP} & 13 & 44.9 & 11.02 & 0.805 & 0.6152 & 20.18 & 0.001 & Significant  \\ \hline
		\textbf{EV} & 12 & 28.4 & 2.702 & 0.795 & 0.5947 & 17.14 & 0.002 & Significant  \\ \hline
		\textbf{MR} & 14 & -4.1 & 4.424 & 0.869 & 0.7348 & 37.03 & 0 & Significant  \\ \hline
		\textbf{PL} & 14 & -1.63 & 1.754 & 0.869 & 0.7349 & 37.03 & 0 & Significant  \\ \hline
		\textbf{ST} & 14 & 53.4 & 0.235 & 0.139 & 0 & 0.24 & 0.635 & Not significant  \\ \hline
		\textbf{MV} & 14 & 0.5 & 11.94 & 0.847 & 0.6936 & 30.43 & 0 & Significant  \\ \hline
	\end{tabular}
\end{table}
	\clearpage
\begin{table}[!ht]
	\centering
	\caption{Statistical parameters for the  linear QSPR model for $ES(G)$}
	\label{table12}
	\begin{tabular}{|l|l|l|l|l|l|l|l|l|}
		\hline
		\textbf{} & \textbf{$N$} & \textbf{$a$} & \textbf{$b$} & \textbf{$r$} & \textbf{$R^2(ad\!j)$} & \textbf{$F$-value} & \textbf{$p$-value } & \textbf{Indicator} \\ \hline
		\textbf{BP} & 13 & 351 & 0.483 & 0.454 & 0.1337 & 2.85 & 0.119 & Not significant  \\ \hline
		\textbf{MP} & 14 & 154.3 & 0.0748 & 0.421 & 10.9 & 2.59 & 0.133 & Not significant  \\ \hline
		\textbf{FP} & 13 & 213 & 0.211 & 0.341 & 0.0362 & 1.45 & 0.254 & Not significant  \\ \hline
		\textbf{EV} & 12 & 76.7 & 0.0398 & 0.269 & 0 & 0.78 & 0.399 & Not significant  \\ \hline
		\textbf{MR} & 14 & 61.2 & 0.0875 & 0.378 & 0.0715 & 2 & 0.183 & Not significant  \\ \hline
		\textbf{PL} & 14 & 24.3 & 0.0347 & 0.378 & 0.0715 & 2 & 0.183 & Not significant  \\ \hline
		\textbf{ST} & 14 & 53.65 & 0.0129 & 0.169 & 0 & 0.35 & 0.565 & Not significant  \\ \hline
		\textbf{MV} & 14 & 185.9 & 0.212 & 0.331 & 0.0357 & 1.48 & 0.247 & Not significant  \\ \hline
	\end{tabular}
\end{table}

\begin{table}[!ht]
	\centering
	\caption{Statistical parameters for the  linear QSPR model for $EG(G)$}
	\label{table13}
	\begin{tabular}{|l|l|l|l|l|l|l|l|l|}
		\hline
		\textbf{} & \textbf{$N$} & \textbf{$a$} & \textbf{$b$} & \textbf{$r$} & \textbf{$R^2(ad\!j)$} & \textbf{$F$-value} & \textbf{$p$-value } & \textbf{Indicator} \\ \hline
		\textbf{BP} & 13 & 135.5 & 0.579 & 0.816 & 0.6358 & 21.95 & 0.001 & Significant  \\ \hline
		\textbf{MP} & 14 & 145.6 & 0.0549 & 0.466 & 0.1515 & 0.0332 & 0.093 & Not significant  \\ \hline
		\textbf{FP} & 13 & 77.4 & 0.3125 & 0.756 & 0.5332 & 14.71 & 0.003 & Significant  \\ \hline
		\textbf{EV} & 12 & 41.6 & 0.07 & 0.69 & 0.4231 & 9.07 & 0.013 & Significant  \\ \hline
		\textbf{MR} & 14 & -0.6 & 0.1387 & 0.902 & 0.7982 & 52.42 & 0 & Significant  \\ \hline
		\textbf{PL} & 14 & -0.25 & 0.05502 & 0.902 & 0.7981 & 52.4 & 0 & Significant  \\ \hline
		\textbf{ST} & 14 & 58.8 & -0.0001 & -0.002 & 0 & 0 & 0.995 & Not significant  \\ \hline
		\textbf{MV} & 14 & 12.2 & 0.3712 & 0.871 & 0.7395 & 37.9 & 0 & Significant  \\ \hline
	\end{tabular}
\end{table}

	This section has been focused on the calculation of Regression parameters where the sample size, constant or intercept of $y$, slope, and the correlation coefficient are indicated by $N, A, b,$ and $r$ respectively. $r^2$ gives the variability percentage of response variable which is explicated by the linear model. Influence of each term tests the null hypothesis which indicates zero coefficients, by using its
	$p$-value. Conversely, a higher (insignificant) $p$-value indicates that changes in the predictor are not related with changes in the response. Assume we are doing a test where the null hypothesis states that all of the regression coefficients are equal to zero. This test proceeds evaluating the models with no explanatory variables and then measures the enhancement of the model with additive coefficients, by providing an $F$ value as the outcome. Somehow, in this scenario, model is unable to provide precise predictions. The Tables \ref{table7} - \ref{table13} show the statistical parameters of linear QSPR models for a range of topological indices. Each model gives a significant result, where the $p$ value is lower or equal to the significance level $\left( 0.05\right) $.
		
	Tuberculosis (TB) is a bacterial infection caused by the Mycobacterium tuberculosis bacteria. This affects the proper functioning of lungs, while threatening the survival of human beings by diffusing to other body parts, specially brain and spine.
	
	A molecular descriptor is a mathematical formula that can be used on any graph representing a molecule structure. Specifically, a chemical structure is represented by a singular numerical value. Graph theory identifies this numerical value as a topological descriptor, which can be correlated with a molecular element and form a molecular index or specifically a topological index (TI). Topology of the graph is used to describe by a topology index which is a result of the transformation of a molecular graph, where calculating these indices could assist in further examining some sort of physio-chemical molecular aspects.
	
	By calculating these topological indices, it is feasible to assess mathematical values and conduct additional investigation into certain physiochemical aspects of a molecule. Topological indices are created by transforming a molecular graph into a numerical value that represents the graph's topology.
	Molecular topological indices have a crucial function in mathematical chemistry, particularly in the study of quantitative structure-property relationship (QSPR) and quantitative structure-activity relationship (QSAR).
	
	Table \ref{table2} consists with values of physical properties and Table \ref{table1} consists with degree based  topological indices for the proposed technique, while Table \ref{table5}  and Figure 3 indicate their correlation. Table \ref{table3} consists of the values of distance based indices for the traditional method while Table  \ref{table4} illustrates the novel distance/ degree distance based topological indices for the proposed method. Also, Table \ref{table6} gives the the correlations for novel distance/degree distance based indices. Upon examining correlation coefficients horizontally for physical properties under consideration, we see that \(W(G)\) gives highest correlation coefficient for  Molar refraction \((r=0.918)\) and polarizability \((r=0.918)\). \(HW(G)\) gives highest correlation coefficient for molar refraction and Polarizability \((r=0.937)\). \(WW(G)\) gives highest correlation coefficient for the molar refraction \(r=0.931\), polarizability \((r=0.931)\). \(Ha(G)\) gives highest correlation coefficient for boiling point \((r=0.785)\). \(BB(G)\) gives highest correlation coefficient for boiling point \(r=0.888\). \(S\!ch(G)\) gives highest correlation coefficient for boiling point \((r=0.454)\). \(Gut(G)\) gives highest correlation coefficient for molar refraction, polarizability\((r=0.902)\). When we look vertically, boiling point has also good correlation with \(HW(G), WW(G)\), i.e., \(r=0.912\) and \((r=0.910)\) respectively. Flash point has good correlation with \(WW(G)\),i.e.,  \(r=0.843\).  Molar refraction has also good correlation with \(HW(G)\) and \(WW(G)\), i.e., \(r=0.937\) and \(r=0.931\). Polarizability has also good correlation with \(HW(G)\) and \(WW(G)\), i.e., \((r=0.937)\) , \(r=0.931\). Molar volume has also good correlation with \(HW(G), WW(G)\), i.e., \(r=0.918, 0.911\) respectively.

	Table \ref{table7} - \ref{table13} show different statistical parameters of correlation between values of five distance based and two degree-distance based topological indices and eight physical properties of medicine. The main drawback of our study is the lack of satisfactory correlate findings among degree-based topological index, and the melting point and surface tension of anti-tuberculosis drugs. However, the theoretical analysis of our study could strengthen the prediction of anti-tuberculosis medicinal attributes with zero involvement of experimentation.
	
	Figure 4 shows the comparison of degree based indices for traditional method. Here the values of topological indices vary upto 1600. Also, Figure 5 shows the comparison of the existing degree based topological indices for the proposed method and here the values of topological indices vary within the range \(0-4000\). Next, Figure 6 illustrates the comparison of distance based/degree distance topological indices for proposed method while the value of the indices vary within the range \(0-1200\).
	
	\section*{Conclusions}
	The current study investigates the drugs that are employed to treat tuberculosis. For each drug, the study employs a variety of degree-based numerical descriptors, distance-based topological indices, and newly introduced distance/degree distance-based indices. A total of 14 anti-tuberculosis pharmaceuticals have been investigated using 10 degree-based indices, 6 distance-based indices, and 6 novel distance/degree distance-based indices.
	
	Here we introduced the method to compute the topological indices. When the computing the topological indices, we considered the three dimensional molecular structures of the relevant medications.  So, Using parameters of the three dimensional chemical graph to build topological indices, those indices capture more information than the respective two dimensional chemical graphs. Also, in the proposed method, we haven't omitted the hydrogen atoms and considered the structures with all the hydrogen atoms. Also, multi bonds of the structures are considered as multi edges of the three dimensional chemical graphs.
	 
	 This study investigated the existing 10 degree based topological indices applying the using the aforesaid points. We suggest that using the proposed method to compute degree based numerical descriptors should give the more reliable results with minimum assumptions when compared to existing computing techniques in cheminformatics.

	Also, the novel 6 distance/ degree distance topological indices were introduced using the actual bond length of each edges as a parameter of indices. Using the actual bond length for the topological index, we suggested that the index should give the more information than using the degree of vertices and/or graph theoretical lengths. 
	
	These indicators may serve as predictive variables in QSPR studies. The physicochemical properties of antituberculosis medicines are examined to evaluate the predictive efficacy of these parameters. The results indicate a highly significant correlation, demonstrating a robust positive linear association among molecular volume, boiling point, enthalpy, flash point, molar refractivity, vapor pressure, and TIs.

	 Research suggests that this theoretical analysis could assist chemists and individuals in the pharmaceutical sector in predicting the features of anti-tuberculosis medications without the need for experimentation. It is conceivable that varying combinations of these medications could be employed to treat distinct illnesses, contingent upon the scope of the topological indices calculated in this study. In this study, we have determined the correlation coefficient for several topological indices. This finding will assist chemists in developing novel medications by combining drugs that are highly positively correlated.
	
	\section*{Future works}

	As a future work, studying the connection between the molecular structure and physical characteristics of drugs provides valuable knowledge for predicting the effectiveness and safety of new treatments. Researchers can expedite the pharmaceutical creation process by accurately associating innovative topological indices with numerous physical features, such as solubility, bio-availability, and stability. The objective of this approach is to provide a resilient structure for assessing prospective drug candidates throughout the first stages of their development, with the aim of improving effectiveness and minimizing expenses in pharmaceutical research. Also, numerical methods can be used to obtain best fitted curve (an interpolated curve) of topological index and the relevant physical property.

\end{document}